\magnification=1200

\def\d{\delta}\def\e{\epsilon}

\def\k{\kappa}\def\l{\lambda}\def\s{\sigma}\def\t{\tau}
\def\y{\eta}

\def\de{\partial}
\def\id{\equiv}\def\ha{{1\over 2}}

\def\tran{transformations }

\def\pb{Poisson brackets }

\def\poi{Poincar\'e }

\def\tl{transformation law }
\def\wrt{with respect to }
\def\eom{equations of motion }
\def\cr{commutation relations }

\def\section#1{\bigskip\noindent{\bf#1}\smallskip}

\def\PL#1{Phys.\ Lett.\ {\bf#1}}
\def\PRL#1{Phys.\ Rev.\ Lett.\ {\bf#1}}
\def\PR#1{Phys.\ Rev.\ {\bf#1}}

 \def\IJMP#1{Int.\ J. Mod.\ Phys.\ {\bf #1}}
\def\MPL#1{Mod.\ Phys.\ Lett.\ {\bf #1}}

\def\ref#1{\medskip\everypar={\hangindent 2\parindent}#1}
\def\beginref{\begingroup
\bigskip
\centerline{\bf References}
\nobreak\noindent}
\def\endref{\par\endgroup}

\def\emp{e^{-2p_0/\k}}\def\ump{\left(1-{p_0\over\k}\right)}
\def\umm{{\k^2-m^2\over\k^2}}


{\nopagenumbers
\line{\hfil April 2007}
\vskip80pt
\centerline{\bf Doubly Special Relativity and Finsler geometry}
\vskip40pt
\centerline{
{\bf S. Mignemi}\footnote{$^\ddagger$}{e-mail:
smignemi@unica.it}}
\vskip10pt
\centerline {Dipartimento di Matematica, Universit\`a di Cagliari}
\centerline{viale Merello 92, 09123 Cagliari, Italy}
\centerline{and INFN, Sezione di Cagliari}
\vskip100pt
\centerline{\bf Abstract}
\vskip10pt
{\noindent
We discuss the recent proposal of implementing Doubly Special Relativity in configuration
space by means of Finsler geometry. Although this formalism leads to a consistent description
of the dynamics of a particle, it does not seem to give a complete description of the physics.
In particular, the Finsler line element is not invariant under the deformed Lorentz
transformations of Doubly Special Relativity.
We study in detail some simple applications of the formalism.
}
\vskip100pt\
\vfil\eject}

\section{1. Introduction}
In a recent paper [1] a relation between modified dispersion relations and Finsler geometry
has been proposed, and in particular it has been observed that Doubly (or Deformed) Special
Relativity (DSR) [2-4] can be realized in ordinary (commutative) configuration space as a
(mass-dependent) Finsler geometry.

We recall that Finsler geometry is a generalization of Riemann geometry whose
metric can depend both on position and velocity [5].
DSR models instead postulate a deformation of the standard \poi invariance of
special relativity such that the momenta transform non-linearly under boosts, leaving
invariant a fundamental energy scale $\k$ (usually identified with the Planck energy).
Such deformation can be obtained by suitably modifying the generators of boosts,
and is not unique.
The implementation of DSR in configuration space is not obvious, and has been extensively
debated [6-9]. It is clear, however, that coordinate transformations consistent with DSR must
be momentum dependent,
and hence the possibility of a momentum (or velocity)-dependent geometry emerges.

This line of thought has been pursued in ref.\ [1]. However, although the formalism proposed
there yields geodesics equations that transform covariantly
under the deformed Lorentz transformations (DLT) of DSR, the Finsler line element is not
invariant.
As a consequence, it is not possible to define an invariant separation between events
and it is difficult to identify a physical proper time in this framework.
Moreover, it does not seem that the Finsler line element be the most natural affine parameter
in this formalism.

In the present paper, we discuss these difficulties and give some explicit examples in the case
of the best known DSR models.

\bigskip
DLT are a deformation of \poi algebra that do not alter the \cr between rotation
generators $M_i$ and boosts generators $N_i$,\footnote{*}{We denote spacetime indices
by $a,b,\dots$ and spatial indices by $i, j,\dots$.}
$$\{M_i,M_j\}=\e_{ijk}M_k,\qquad\{M_i,N_j\}=\e_{ijk}N_k ,\qquad\{N_i,N_j\}=-\e_{ijk}M_k,
\eqno(1.1)$$
and between rotations and translations generators $p_a$,
$$\{M_i,p_0\}=0,\qquad\{M_i,p_j\}=\e_{ijk}p_k,\eqno(1.2)$$
but modify the \cr between boosts and translations generators:
$$\{N_i,p_a\}=w^i_a(p),\eqno(1.3)$$
where the $w_a^i$ are nonlinear functions of the momentum $p$ and a dimensional parameter $\k$,
usually identified with the Planck energy.
The dispersion relation of the momentum are defined by means of the Casimir invariant $C(p)$
of the deformed algebra,
$$C(p)=m^2,\eqno(1.4)$$
with $m$ the mass of the particle.

According to [1], the Hamilton equations for a free particle in canonical phase space
can be obtained from the reparametrization-invariant action
$$I=\int\left[\dot q_ap_a-{\l\over2m}(C(p)-m^2)\right]d\t,\eqno(1.5)$$
where $q$ are the coordinates of the particle, $\t$ is an evolution parameter which is
assumed to be invariant under DLT, and a dot denotes derivative with respect to it.
The Lagrange multiplier $\l$ enforces the dispersion relation (1.4), and
the summations are performed \wrt the flat metric $\y_{ab}= (1,-1,-1,-1)$.
Varying with respect to $q_a$ and $p_a$, one obtains
$$\dot q_a={\l\over m}{\de C\over\de p_a},\qquad\dot p_a=0.\eqno(1.6)$$

To write the action in configuration space, one can invert the first equation,
obtaining $p_a=p_a(m\dot q/\l)$. Then, substituting in the dispersion relation, one can
obtain $\l$ as a homogeneous function of $\dot q_a$ of degree one.
Finally, substituting back in the action,
$$I=\int\dot q_ap_a(\dot q)d\t\id \int L(\dot q)d\t.\eqno(1.7)$$
The Lagrangian $L$ depends only on the 4-velocity $\dot q$ and is homogeneous of degree
one in $\dot q$.
One can therefore identify $L$ with a Finsler norm.
The Finsler line element is then defined as [5]
$$ds^2=\ha{\de^2 L^2\over\de q_a\de q_b}\ dq_adq_b,\eqno(1.8)$$
and satisfies
$$ds=Ld\t,\eqno(1.9)$$
by virtue of the homogeneity property of the Lagrangian.

It must be noted however that $L$, and hence $ds$, is not invariant under DLT. In
fact, from the Jacobi identities one can deduce the infinitesimal \tl of the
coordinates $q_a$ [7]. Indeed,
$$\{\{N_i,q_a\},p_b\}+\{\{p_b,N_i\},q_a\}+\{\{q_a,p_b\},N_i\}=0.\eqno(1.10)$$
Assuming canonical \pb between phase space variables,
$$\{q_a,q_b\}=0,\qquad\{p_a,p_b\}=0,\qquad\{q_a,p_b\}=\y_{ab},\eqno(1.11)$$
the last term of (1.10) vanishes, and one gets after integration
$$\{N_i,q_a\}=-{\de w^i_b\over\de p_a}\ q_b.\eqno(1.12)$$

It is then evident that the variation of the Finsler norm does not vanish in general.
We can write in fact
$$\d_i L(\dot q)=\d_i(\dot q_ap_a)={d\over d\t}\left[\d_i(q_ap_a)\right]-\d_i(q_a\dot p_a)=
{d\over d\t}\left(w_a^iq_a-p_b{\de w_a^i\over\de p_b}\ q_a\right),\eqno(1.13)$$
where the term $\d_i(q_a\dot p_a)$ vanishes as a consequence of (1.3) and (1.12).
The remaining term can vanish only if the $w^i_a$ are homogeneous function of degree one of
the $p$, which rules out the standard DSR models, where a dimensionful parameter $\k$ enters
in the definition of the $w^i_a$. Therefore, also the line element $ds$ is not
invariant in general.
The covariance of the geodesics equations obtained by varying (1.7) however persists since
$\d_iL$ is a total derivative.

One may notice that in the present formalism the natural definition of affine parameter
seems not to coincide with the Finsler line element $ds=Ld\t$, but rather with $d\s\id\l d\t$.
In terms of $\s$, the Hamilton equations can in fact be written in the usual form
$${dq_a\over d\s}\id q'_a={1\over m}{\de C\over\de p_a},\qquad {dp_a\over d\s}\id p'_a=0.
\eqno(1.14)$$

Moreover, being $\l$ a homogeneous function $f(\dot q)$ of degree one, one has
$$\l=f(\dot q)=\l f(q'),\eqno(1.15)$$
from which one obtains a constraint on the four-velocity expressed in terms of the
proper time, $f(q')=1$, analogous to the relation $q'^2_0-q'^2_i=1$ of special relativity.

The corresponding equations in terms of the Finsler line element take a much more involved form.
Unfortunately, however, in general also the affine parameter $\s$ is not invariant
under DLT.

\section{2. The Magueijo-Smolin model}
In order to better understand the implications of the previous considerations, it is useful to
consider some simple examples.
The simplest one is the Magueijo-Smolin (MS) model [4], whose Lagrangian formulation has been
studied in [10].
The dispersion relation is
$${p_0^2-p_i^2\over(1-p_0/\k)^2}=m^2.\eqno(2.1)$$
which is left invariant by the DLT
$$\{N_i,p_0\}=p_i\ump,\qquad\{N_i,p_j\}=\d_{ij}p_0-{1\over\k}p_ip_j.\eqno(2.2)$$
From (1.10) follows [7]
$$\{N_i,q_0\}=q_i+{p_i\over\k}q_0,\qquad\{N_i,q_j\}=\ump q_0\d_{ij}+
{1\over\k}(p_kq_k\d_{ij}+p_iq_j).\eqno(2.3)$$
As already noted in [4], the deformed algebra is generated by the standard rotation generators,
while the boost generators are given by
$$N_i=q_0p_i-q_ip_0-{p_i\over\k}\ q_ap_a.\eqno(2.4)$$
In order to simplify the calculation of the action, it is useful to write it as [10]
$$I=\int\left[\dot q_ap_a-{\l\over2m}\left(p_a^2-m^2\ump^2\right)\right]d\t.\eqno(2.5)$$
The field equations then read
$$\dot q_0={\l\over m}\left(\umm p_0+{m^2\over\k}\right),\qquad\dot q_i={\l\over m}p_i,\eqno(2.6)$$
and can be inverted explicitly,
$$p_0={\k^2\,m\over \k^2-m^2}\left({\dot q_0\over\l}-{m\over\k}\right),\qquad p_i=m{\dot q_i\over\l}.
\eqno(2.7)$$
Substituting into (2.1) one can obtain $\l$ as a function of $\dot q$,
$$\l=\sqrt{\dot q_0^2-\umm\ \dot q_i^2}.\eqno(2.8)$$

The action can then be written in terms of the $\dot q$ as
$$I=\int\dot q_ap_a(\dot q)d\t={\k^2m\over\k^2-m^2}\int\left(\l-{m\over\k}\,\dot q_0\right)d\t.
\eqno(2.9)$$
Notice that the last term is a total derivative and does not contribute to the field equations,
that read
$${d\over d\t}\left(\dot q_a\over\l\right)=0,\eqno(2.10)$$
or, in terms of the affine parameter $\s$ defined in section 1, $q_a''=0$, as in special relativity.
However, now the 4-velocity satisfies the constraint
$$q'^2_0-\umm\ q'^2_i=1.\eqno(2.11)$$
In terms of the parameter $\s$, the Lagrangian reads
$$L={\k^2m\over\k^2-m^2}\ \l\left(1-{m\over\k}\ q_0'\right).\eqno(2.12)$$
Using the Finsler parameter $s$ the equations would take a much more involved form.

We want now to investigate the transformation properties of the action. Under the action of a boost
$N_i$, by (2.2)-(2.3),
$$\d_i L=\d_i(\dot q_ap_a)={1\over\k}{d\over d\t}(p_ip_aq_a).\eqno(2.13)$$
In order to calculate the variation of $\l$, we consider for simplicity the two-dimensional case,
where one has a single boost generator and the \tran (2.2)-(2.3) reduce to:
$$\d p_0=p_1-{p_0p_1\over\k},\qquad\d p_1=p_0-{p_1^2\over\k},\eqno(2.14)$$
$$\d q_0=q_1+{p_1\over\k}q_0,\qquad\d q_1=q_0-{p_0\over\k}q_0+{p_1\over\k}q_1.\eqno(2.15)$$
Under these transformations, also the variation of $\l$ reduces to a total derivative,
$$\d\l={1\over\k}{d\over d\t}\left[q_1+{p_1\over\k}q_0+{\k^2-m^2\over\k^2m^2}\,p_1(p_aq_a)\right].
\eqno(2.16)$$
On shell, the variation of $L$ and $\l$ take the simpler form
$$\d L={\k\,m^2\dot q_1\over\k^2-m^2}\left(1-{m\dot q_0\over\k\l}\right),
\qquad\d\l=-{2m\dot q_1\over\k}.\eqno(2.17)$$
Thus, although the \eom are covariant under the DLT, neither the lagrangian $L$ nor the Lagrange
multiplier $\l$ are invariant, but their variation is a total derivative.
This of course is a problem if we wish to define an invariant proper time.

\section{3. The Lukierski-Nowicki-Ruegg model}
Another important example is given by the $\k$-\poi algebra of Lukierski-Nowicki-Ruegg [3].
In this case
$$\{N_i,p_0\}=p_i,\qquad\{N_i,p_j\}={\k\over2}\left(1-\emp\right)\d_{ij}+
{1\over2\k}(p_kp_k\d_{ij}-2p_ip_j),\eqno(3.1)$$
and hence [7]
$$\{N_i,q_0\}=\emp q_i,\qquad\{N_i,q_j\}=q_0\d_{ij}+{1\over\k}(p_kq_k\d_{ij}+p_iq_j-p_jq_i).
\eqno(3.2)$$
The dispersion relation invariant under (3.1)-(3.2) is
$$C(p)=4\k^2\sinh^2(p_0/2\k)-e^{p_0/\k}p_i^2=m^2\eqno(3.3)$$
It is then easy to write down the deformed generators of the boosts
$$N_i=p_iq_0-{\k\over2}\left(1-\emp\right)q_i-{1\over2\k}(p_kp_kq_i-2p_ip_kq_k).\eqno(3.4)$$

Varying the action (1.4) one obtains the Hamilton equations
$$\dot q_0={\l\over m}\left[\sinh(p_0/\k)-e^{p_0/\k}{p_i^2\over2\k}\right],\qquad
\dot q_i={\l\over m}\ e^{p_0/\k}p_i.\eqno(3.5)$$
In this case it is not possible to invert explicitly the equations for $p$ in terms of $\dot q$,
so we expand them in powers of $m/\k$,
$$p_0\sim m\left({\dot q_0\over\l}+{m\over2\k}\ {\dot q_i^2\over\l^2}\right),\qquad
p_i\sim m\left({\dot q_i\over\l}-{m\over\k}\ {\dot q_0\dot q_i\over\l^2}\right).\eqno(3.6)$$
Substituting in (3.3),
$$\l\sim\sqrt{\dot q_0^2-\dot q_i^2}+{m\over\k}\ {\dot q_0\dot q_i^2\over
\dot q_0^2-\dot q_i^2},\eqno(3.7)$$
and finally,
$$I\sim m\int\left[\sqrt{\dot q_0^2-\dot q_i^2}+{m\over2\k}\ {\dot q_0\dot q_i^2\over
\dot q_0^2-\dot q_i^2}\right]d\t.\eqno(3.8)$$
In terms of the affine parameter $\s$,
$$L\sim m\l\left(1-{m\over2\k}\,q'_0q'^2_i\right).\eqno(3.9)$$
Note that, contrary to the MS case, now $L$ and $\l$ do not differ by a total derivative.

To consider the effect of the \tran (3.1)-(3.2) on the Lagrangian, we again consider the
two-dimensional case, for which
$$\d p_0=p_1,\qquad\d p_1={\k\over2}\left(1-\emp\right)+{p_1^2\over2\k}\eqno(3.10)$$
$$\d q_0=\emp q_1,\qquad\d q_1=q_0+{p_1\over\k}q_1.
\eqno(3.11)$$
Under these transformations, the Lagrangian changes by a total derivative (see 2.13),
$$\d L={d\over dt}\left[\left(p_0-{\k\over2}(1-e^{-2p_0/\k})-{p_1^2\over2\k}\right)q_1\right]
\sim-{1\over2\k}{d\over d\t}\left[(2p_0^2+p_1^2)q_1\right].\eqno(3.12)$$
The transformation properties of $\l$ are instead quite involved and its variation
under DLT is not even a total derivative.

\section{4. Conclusions}
As we have shown, the formalism of ref.\ [1] permits to write down DLT-covariant
equations for the geodesic motion of a point particle in canonical configuration space
of DSR.
However, it does not allow to define an invariant affine parameter, and hence to identify
the physical proper time.

The situation does not change by passing to noncommutative spacetime coordinates. In fact
in this case, in order to obtain the correct Hamilton equations, one has to modify the
term $p\dot q$ in the action [7], and the new term is still not invariant under the DLT
compatible with the new symplectic structure.

It seems therefore that although Finsler spaces are useful for studying the motion of
DSR particles, they do not catch the full structure of the theory.
This is presumably due to the fact that Finsler spaces are tailored for the study of
homogeneous dispersion relations, while DSR dispersion relations cannot be homogeneous
because of the presence of a dimensional constant $\k$. Moreover, the formalism of Finsler
spaces is built in such a way to avoid the presence of Lagrangian multipliers in the action
principle [11].

A possible solution might be to consider some generalizations of Finsler spaces, either
defining a metric structure in the full phase space, instead of configuration space, or
perhaps by relaxing the requirement of homogeneity of the Finsler metric. A last possibility
is that the formalism work better when a five-dimensional configuration space is
considered as in [9,12].

\section{Acknowledgments}
{\noindent I wish to thank Stefano Liberati for a useful discussion.}
\bigskip
\beginref
\ref [1] F. Girelli, S. Liberati and L. Sindoni, \PR{D75}, 064015 (2007).
\ref [2] G. Amelino-Camelia, \IJMP{D11}, 35 (2002), \PL{B510}, 255 (2001).
\ref [3] J. Lukierski, A. Nowicki, H. Ruegg and V.N. Tolstoy, \PL{B264}, 331 (1991);
J. Lukierski, A. Nowicki and H. Ruegg, \PL{B293}, 344 (1992).
\ref [4] J. Magueijo and L. Smolin, \PRL{88}, 190403 (2002).
\ref [5] H. Rund {\it The differential geometry of Finsler spaces}, Springer-Verlag
1959.
\ref [6] J. Kowalski-Glikman, \MPL{A17}, 1 (2002).
\ref [7] S. Mignemi, \PR{D68}, 065029 (2003); \PR{D72}, 087703 (2005).
\ref [8] D. Kimberly, J. Magueijo and J. Medeiros, \PR{D70}, 084007 (2004).
\ref [9] A.A. Deriglazov and B.F. Rizzuti, \PR{D71}, 123515 (2005).
\ref [10] S. Ghosh, \PR{D74}, 084019 (2006).
\ref [11] H. Rund {\it The Hamilton-Jacobi theory in the calculus of variations}, Krieger
1973.
\ref [12] F. Girelli, T. Konopka, J. Kowalski-Glikman and E.R. Livine, \PR{D73}, 045009 (2006).
\endref
\end